# Neural Underpinnings of Decoupled Ethical Behavior in Adolescents as an Interaction of Peer and Personal Values

Manvi Jain[1], Karsheet Negi[2], Pooja S. Sahni[1], Mannu Brahmi[1], Neha Singh,[3] Dayal Pyari Srivastava,[3] Jyoti Kumar[1]

[1]Department of Design, Indian Institute of Technology, Delhi, 110116, India
[2]Department of Design, Delhi Technological University, Delhi, 110042, India
[3]Department of Physics and Computer Science, Dayalbagh Educational Institute, Agra, 282004, India

**Abstract**

In the present study an attempt has been made to understand how peer unethical behavior may decouple personal ethical behaviour in adolescents. An interactive game was developed that presented the player with two situations either of frustration or gratitude for the partner in game. Personal ethics was measured through Personal Values Questionnaire. The responses (reward sharing) given by participants were recorded as well as neural signals using an EEG. There was significant correlation found between personal ethics and reward sharing. Preliminary analysis was focused on studying the differences in lower brain frequencies (0.1-4Hz) when the participants developed frustration, in contrast to when they experienced gratitude for the partner. The study presents three case studies in which delta frequencies increased in cases when frustration was experienced and decreased when gratitude was experienced. The study focused on understanding the neural underpinnings of corresponding modified behavior in adolescents. The findings highlight an increase in delta frequencies when apparent frustration was developed in adolescents due to their peer unethical behavior. The delta frequencies lowered when participants were tested for ethical behavior. The results concluded that based on personal value types, adolescents tend to develop frustration toward perceived unethical behavior and carry it over to other unrelated peers. This study is highly explorative in nature, with preliminary analysis using only three case studies, having a small sample size. However, this study introduces new methods to social cognition in adolescents..

## Introduction

Ethical Behavior in Adolescents: Some theories state that morality becomes a part of an individual's self-concept during adolescence (Colby and Damon 1992; Hardy and Carlo 2005; Moshman n.d.) forming one's moral identity. Strong moral identity engages individuals in moral actions as it drives them toward a sense of obligation to behave in ways that align with their moral values (Hardy and Carlo 2005; Blasi 1983). Moral judgment in adolescents contributes majorly to morally relevant actions that may include responsibility towards society, prosocial acts, moral actions toward peers, etc. (Hart, Atkins, and Donnelly, n.d.; Hertz and Krettenauer 2016).

Role of Personal values in Ethical Behavior: From a social-cognitive perspective, moral identity is a cognitive representation of moral values, goals, traits and behavioral scripts (Pohling et al. 2016; Hannah, Avolio, and May

2011). Therefore, personal values and personality can be identified as specific aspects of moral behavior. Values are used to characterize cultural groups, societies, and individuals, to trace change over time, and to explain the motivational bases of attitudes and behavior. Recent theoretical and methodological developments (Schwartz 1992; Berry et al. 1997) have brought about a resurgence of research on values. Schwartz's theory of values identifies ten distinct types of values and describes the dynamic relations among them. This list of values contains some contrasting values (e.g., benevolence and power) whereas some are compatible with each other (e.g., conformity and security) (Schwartz 2012). The circular structure in Figure 1 represents pairs of major contrasting values namely Self-Transcendence and Self-Enhancement that include opposite values. Self-transcendence includes universalism and benevolence which positively correlate with empathy and ethical competence (Hofmann-Towfigh 2007). Self-enhancement, on the other hand, is found to be negatively correlated with moral judgment and discourse competence.

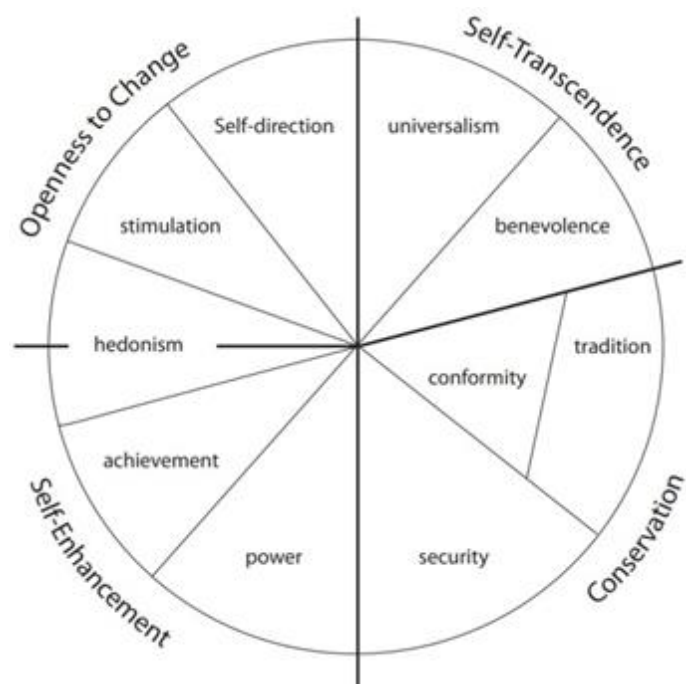

Figure 1: The circular structure of ten major values given in Schwartz's theory of values. It highlights two pairs of major contrasting groups of values including Self-Transcendence and Self-Enhancement that include opposite values.

Social factors in Behavior Change: The value systems are generally stable over time; however they may be prone to change depending upon one's circumstances (Rokeach 1973). According to (Padilla-Walker and Carlo 2014), goals to behave in a moral manner can be motivated by certain internalized (e.g., personal values) and external (e.g., punishment or reward) concerns. At middle and high school years in adolescence, the role of peers becomes critical in affecting prosocial behavior in individuals. It is highly likely that positive feedback from one's peers in the form of social approval and acceptance of their behavior may increase their empathetic, prosocial and moral behavior. In contrast, negative feedback for the behavior that is moral according to the individual, may result in reversing their original behavior. Social influence works at multiple levels, from gradual and permanent modifications in mood, language, gestures, etc., to more immediate influences on an individual's social attitudes and activities (Burnett et al. 2011).

Measures of Behavioral Change: In the past, several studies measured interaction of changes in adolescent behavior with age, group influence etc. A review suggests that behavioral economic paradigms which engage participants in structured competitive or cooperative interactions, reveal subtle differences in the degree of mental perspective-taking (Burnett et al. 2011). Such tasks are quantified as the amount of reward money/tokens exchanged with social partners (Berg, Dickhaut, and McCabe 1995; Binmore 2007). Neuroimaging studies show heightened activity in the reward system of the brain when such a game paradigm is introduced to adults. However, in similar circumstances, peer influence plays an important role in the case of adolescents. As previously mentioned, some studies suggest positive societal influence, on the other hand, some studies (Geier and Luna 2009) report peer influence on potentially harmful behaviors such as increase in risk taking behavior in simulated driving studies, especially in adolescents. The same can be understood from a neuropsychological perspective. The present study uses a novel behavioral game paradigm that is designed to study neural underpinnings of peer influence on adolescent brains. This is implemented by simulating conditions which aim to develop frustration towards peers. Neurophysiologically, frustration is correlated with dominance of lower brain frequencies (delta - 0.1-4Hz and theta - 4-8Hz). Hypothetically, in the present study, if an adolescent develops frustration, they tend to behave in a way that is morally unacceptable and in contrast, if they lack frustration biomarkers, it is assumed that they may behave in a morally correct manner.

In further sections, the methodology used in the study has been discussed where the game paradigm is elaborated. Followed by the methods, there is a results section that discusses the outcomes and finally the discussion section elaborates on inferences drawn on the basis of outcomes.

## Methods

Participants: A cohort of 16 school girls aged 14-16 years (M=14.6years, SD=1.8 years) participated in this study. All the participants were taken from the class 9 of the same school to control any differences due to education type, socio-economic status or other ecological factors. In the present study which is a part of a larger study with the present dataset, only 3 representative case studies have been discussed. The participants signed consent forms before participating and received incentives for their participation in the study.

Game-based task: An interactive game environment was designed consisting of different levels of subitizing tasks (Clements 1999). The task requires the participant to intuitively identify the majority color of the objects given on the screen and press the corresponding button. The level of difficulty ranged from trial to trial (as shown in Figure 2). The task was presented in three blocks of 10 games each. The ten games consisted of 10 rounds each.

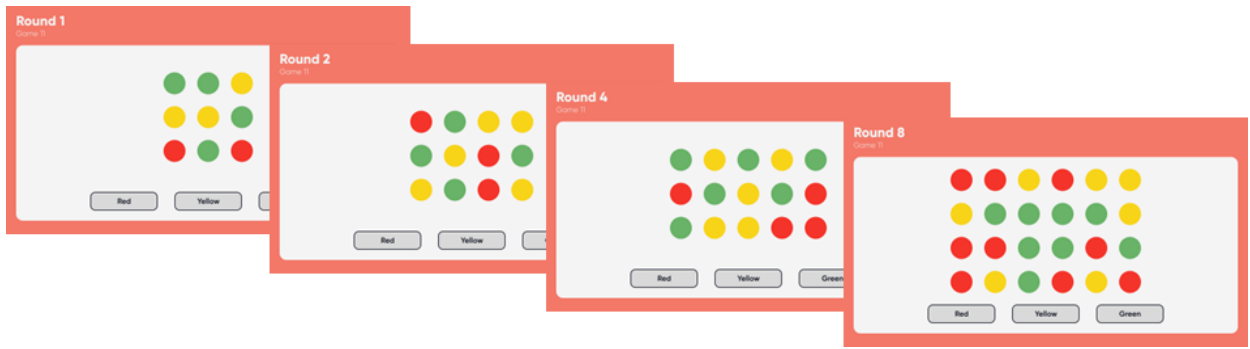

Figure 2: Level of difficulty in Subitizing task: The participant is required to intuitively identify the majority color of circles. The round 1 is the easiest level consisting of 3*3 object matrix, rounds 2,3 are medium level consist of 3*4 object matrix, rounds 4,5,6,7 are low difficulty level consist of 3*5 object matrix and rounds 8,9,10 are high difficulty level consist of 5*6 object matrix.

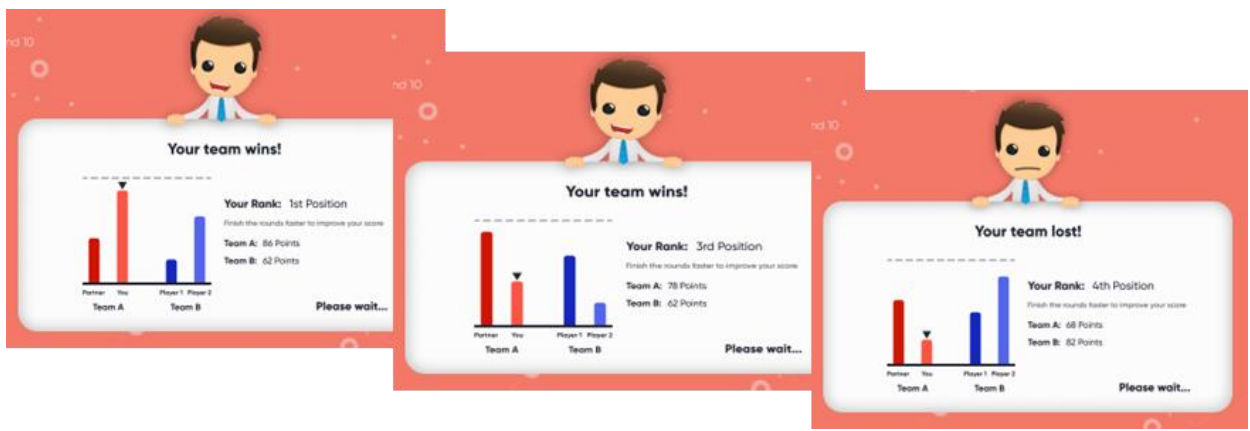

Figure 3: Result screens: The participant received different relative positions in the game.

At the beginning of each block, different virtual partners and opponent teams are introduced and staged as the participant is playing a multiplayer online game with them. However, the responses given by the partner are predefined and there is no virtual partner/team connected in real time. At the end of each game (after every 10 rounds), a result screen was shown to participants which represents the relative position of the participant as compared to their virtual partner and opponent team (Figure 3). The winning team receives 5 token rewards and a team member of the winning team randomly receives the reward to share it with the other member by pressing the corresponding button, e,g., press button 4 to share 4 reward tokens, and so on. The responses of the virtual partner are predetermined in an order-effect based manner in order to evoke emotional responses in participants' brains. For the same purpose, each block was designed differently, the first block was the conditioning block for the participants to instill moral judgment toward their partner by giving 80% reward to the team member who came in first position, the second block was the frustration block in which the virtual partner was depicted to be unethical in sharing rewards when the participant came in first position, and the final block was test of morality block which required the participant to share rewards when the partner came in different positions making the team win or lose correspondingly, to test moral behavior of the participant.

Data collection: *Psychological* - For psychological assessment of personal values of the participants, two scales were used, namely recently revised Portrait Values Questionnaire (PVQ-RR) (Schwartz 2021) and Interpersonal Reactivity Index (IRI) (Davis 1983). Other supporting data was also collected using a demographic scale consisting of questions about average grades in school, family education level, etc.; *Neurophysiological* - While playing the game, the participants were wearing a 64 channel EEG device that collects continuous neural signals. Some event-based markers were introduced in the EEG dataset at specific instances of reward sharing/receiving for further analysis. The sampling frequency of the EEG device is 256 Hz. Bandpass or IIR filters were kept between 0.1 to 45 Hz with a notch filter of 50 Hz. For ocular correction, channels Fp1 and Fp2 were used as EOG channels.

**Results**

The results of this study include a multivariate analysis of three cases that include participant/subject no. 6, 7 and 9. Psychological tests were also given to participants to report for personal values such as empathy, self-transcendence and self-enhancement. A scale called IRI was used to measure scores on components of empathy in participants namely cognitive empathy and affective empathy. The other scale used was PVQ-RR which measured two values in personality namely, self-transcendence which is composed of benevolence and universalism and in contrast, self-enhancement which is composed of hedonism, achievement and power (refer to figure 1). The scores of the three subjects for PVQ-RR psychological scale (Mean=15.33, S.D. =6.8) are as follows: Subject 6 scored 62 on ST and 39 on SE, Subject 7 scored 67 on ST and 54 on SE, Subject 9 scored 53 on ST and 43 on SE. Along with these scales, average grades in school were also measured which were found to be similar for all subjects (Mean= 84.10, SD= 0.8). For IRI, Subject 6 scored 29 on CE and 27 on AE, Subject 7 scored 37 on CE and 38 on AE, Subject 9 scored 36 on CE and 15 on AE. Figure 4 shows scores of the three subjects for PVQ-RR and IRI scales. Along with these scales, average grades in school were also measured for the three subjects which was found to be similar for all (Mean= 84.10, SD= 0.8).

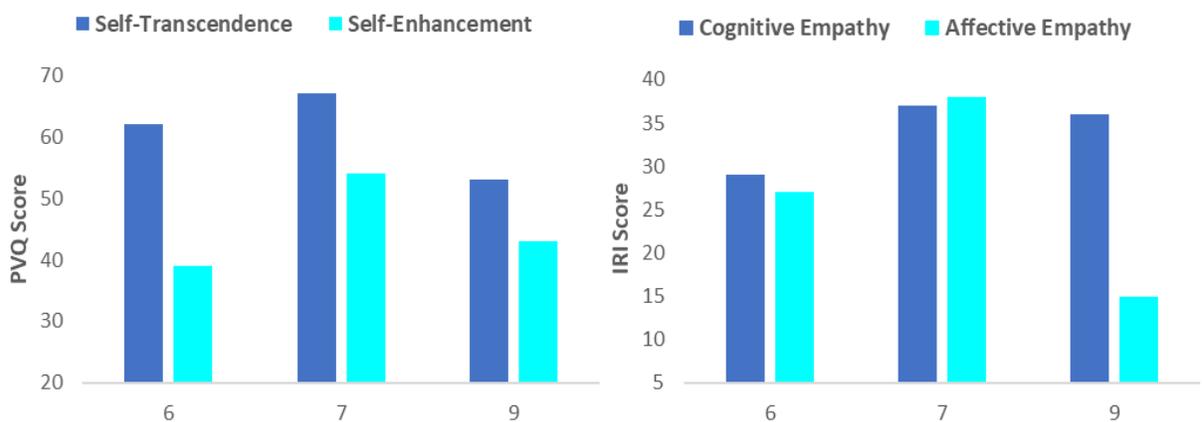

Figure 4. Scores for subject 6, 7 and 9 on PVQ-RR and IRI psychological scales.

The behavioral results are presented in Figure 4 which shows the average number of reward tokens shared by the participants with their apparent virtual partner in the two comparative blocks. The first block i.e., conditioning block is compared with the third block i.e., test of morality block. The average for the two blocks only includes the rounds in which participants came in fourth position and the virtual partner was given first position.

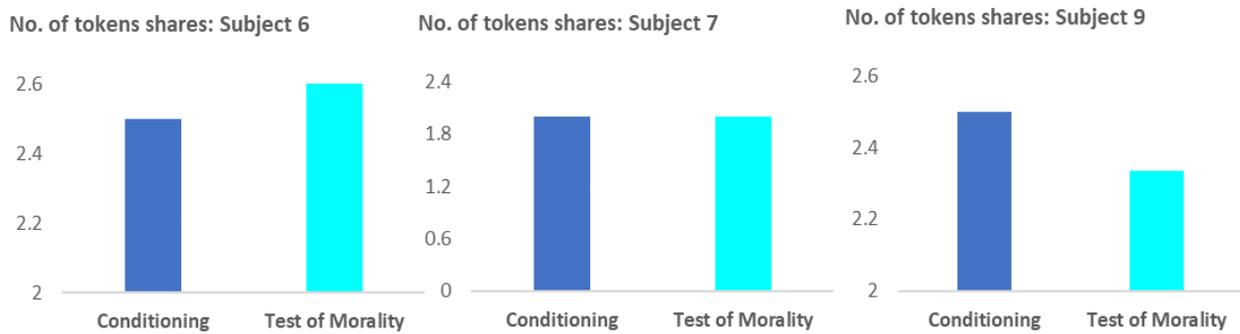

Figure 5: Behavioral results: For subject 6, 7 and 9, the average number of reward tokens (ranging from 1-5), shared by the participant are plotted for the first (conditioning) block and third (test of morality) block.

Neurophysiological analysis was carried out in MATLAB-based software (Brainstorm) using spectral analysis techniques. The specific events of receiving or sharing reward tokens were extracted for an epoch of 1000 milliseconds to observe spectral changes across events. The events with significant differences across conditions and subjects both behaviorally and neurophysiologically were found to be the rounds in which participants came in fourth position and the virtual partner was given first position. Therefore, the EEG signals for the aforementioned events were analyzed. Relative spectral power for different frequency ranges was calculated by subtracting power for second and third blocks from that in the rounds of the first block. The relative power differences across conditions in all subjects were found to be significant for delta frequency range (0.1 to 4Hz) as shown in Figure 5.

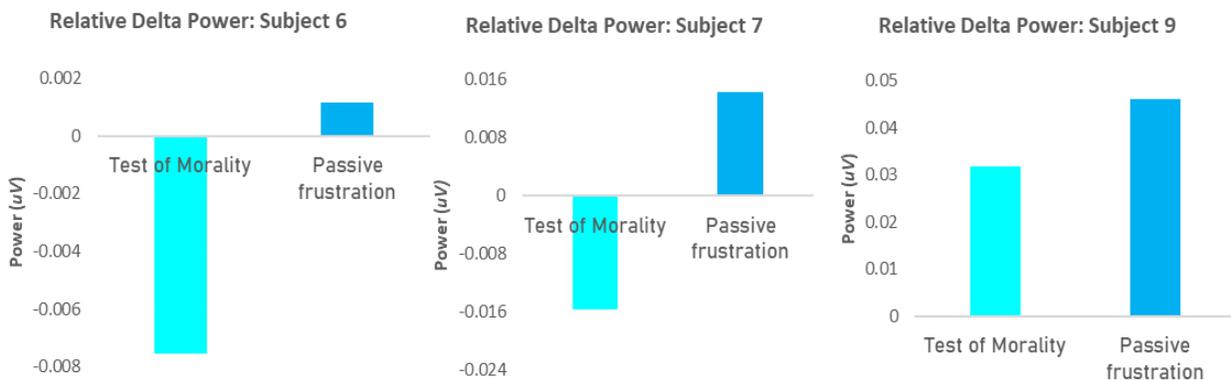

Figure 6: Neurophysiological results: Relative power for delta range (0.1 to 4 Hz) represented for second and third block in the subjects 6, 7 and 9.

**Conclusion**

Case study 1: Subject 6 scored higher on ST than SE with a difference of 23 which appears to be significant; scored equally on empathy representing positive personality traits. Relative delta power for the frustration block does not have significant value whereas for the third block, power has negative value. This depicts that the subject did not develop frustration towards the partner and showed gratitude, also evidenced in Figure 2 having more for reward sharing in the third block.

Case study 2: Subject 7 scored higher on ST than SE with a difference of 13 (mean=15.33) which is not significant; scored equally on empathy representing positive personality traits. Relative delta power for the frustration block has positive value whereas for the third block, power has negative value. This depicts that some level of frustration was developed toward the partner, however it was not carried over in the third block, also evidenced by neutral reward sharing.

Cast study 3 (Subject 9): Subject 9 scored higher on ST than SE with a difference of 10 (mean=15.33) which is not significant. They also scored differently on empathy components, having a higher score for cognitive (36) as compared to affective (15) empathy representing more self-enhancement traits in their personality. Relative delta power for passive frustration blocks has positive value for both blocks. This depicts some level of frustration developed toward the partner, and it was carried over to the next one, also shown by lesser reward sharing in the third block.

Conclusively, the results show that based on personality value types, adolescents tend to develop frustration toward perceived unethical behavior and carry it over to other unrelated peers. This study is highly explorative in nature, with preliminary analysis using only three case studies, having a small sample size. However, the novelty of this study brings about new dimensions to social cognition and personality studies.

**Acknowledgment:** The study has been funded by the VARELA grant offered to author PS. The data collection took place at Dayalbagh Educational Institute, Agra, India where author NS and her supervisor and author DPS contributed in venue and subject procurement as well as data collection. The authors thank all the participants who agreed to join the study.